
\input harvmac
\Title{CERN-TH 6516/92}{Infinite Symmetry in the Quantum Hall Effect}
\centerline{ Andrea CAPPELLI\footnote{*}{\rm
      On leave of absence from I.N.F.N., Firenze, Italy.},
Carlo A. TRUGENBERGER and Guillermo R. ZEMBA}
\bigskip
\centerline{\it Theory Division\footnote{**}{\rm
      Bitnets CAPPELLI, CAT, ZEMBA at CERNVM.}}
\centerline{\it C.E.R.N.}
\centerline{\it 1211 Geneva 23, Switzerland}
\vskip .4in
\centerline{\bf ABSTRACT}
\vskip .2in
Free planar electrons in a uniform magnetic field are shown to possess
the symmetry of area-preserving diffeomorphisms ($W$-infinity
algebra). Intuitively, this is a consequence of gauge invariance,
which forces dynamics to depend only on the flux. The infinity of
generators of this symmetry act within each Landau level, which is
infinite-dimensional in the thermodynamical limit.
The incompressible ground states corresponding to completely
filled Landau levels (integer quantum Hall effect) are shown to
be infinitely symmetric, since they are annihilated by
an infinite subset of generators.
This geometrical characterization of incompressibility
also holds for fractional fillings of the lowest level
(simplest fractional Hall effect) in the presence of Haldane's
effective two-body interactions.
Although these modify the symmetry algebra, the corresponding
incompressible ground states proposed by Laughlin are again symmetric
with respect to the modified infinite algebra.
\vskip .3in
\vfill
\noindent CERN-TH 6516/92

\Date{May 1992}
\newsec{Introduction}

The quantum Hall system
\ref\qhe{R. E. Prange and S. M. Girvin eds., {\it ``The Quantum
Hall effect''}, Springer, New York, (1990).}
is a fascinating example of a {\it quantum fluid}.
Two-dimensional electrons in an external uniform
magnetic field occupy highly degenerate Landau levels. At low
temperatures and large fields, they have strong quantum correlations
which lead to collective motion and macroscopical quantum effects.
The macroscopical quantum states are few, ``simple'' and universal,
and they find their experimental evidence in a discrete series
of quantized values for the Hall conductivity:
\eqn\sig{\eqalign{\sigma_{x y} &=\nu{\displaystyle e^{2} / h} , \cr
\nu &= 1\ (\pm 10^{-8}) ,\ {1\over 3}\ (\pm 10^{-5}) ,\ {1\over 5} ,
\ {1\over 7},\ {2\over 5}, {2\over 7},\dots,\ {6\over 13},\dots,
\ {2 \over 3},\ 2,\ {5 \over 2},\dots,}}
where $\nu$ is interpreted as the {\it filling fraction}, the ratio
between
the number of electrons and the degeneracy of the Landau levels.

The observed universality clearly calls for an approach based on
an effective field theory and renormalization group ideas
\ref\frze{J. Fr\"ohlich and A. Zee, {\it ``Large Scale Physics of the
Quantum Hall Fluid''}, NSF-ITP-91-31 preprint; J. Fr\"ohlich and
T. Kerler, {\it Nucl. Phys.} {\bf B 354} (1991) 369; A. Lopez and
E. Fradkin, Urbana preprint P-92-3-35 (March 1992).}
\ref\lutros{C. A. L\"utken and G. G. Ross, {\it Phys. Rev.} {\bf B 45}
(1992), to appear.}.
However, the most striking feature is the extraordinary precision
of these rational values for the conductivity.
Exact numbers usually
indicate that dynamics is dominated by symmetry and topology, as is
familiar from infinite conformal symmetry in two dimensions
\ref\bpz{A. A. Belavin, A. M. Polyakov and A. B. Zamolodchikov,
{\it Nucl. Phys.} {\bf B 241} (1984) 333.},
and from topological quantization conditions of gauge fields
\ref\curra{S. B. Treiman, R. Jackiw, B. Zumino and E. Witten,
{\it ``Current Algebra and Anomalies''}, Princeton University
Press, New Jersey, (1985).}.
Due to these expectations, the quantum Hall effect has attracted
the interest of theoretical physicists beyond the border of the
solid-state community.

In this paper, we show that an infinite-dimensional
symmetry is indeed present in the Hall system at
integer and (the simplest) fractional fillings. This yields a
strong geometrical characterization of {\it incompressibility}
of the ground state and it may be a key idea to understand the exactness
of the collective behavior of electrons.
Before specifying the content of this paper, let us briefly review
the phenomenology of the problem and previous
approaches based on symmetry principles.

The established phenomenology is based on the seminal work of
Laughlin, Haldane, Halperin and others
\ref\laughc{R. B. Laughlin, {\it Phys. Rev. Lett.} {\bf 50} (1983) 1395.}
\ref\hala{F. D. M. Haldane, {\it Phys. Rev. Lett.} {\bf 51} (1983) 605.}
\ref\halp{B. I. Halperin, {\it Phys. Rev. Lett.} {\bf 52} (1984) 1583.}
\qhe.
It comprises a body of theoretical arguments
substantiated by numerical simulations,
which account for the exactness and stability of the ground
state at the plateaus of the Hall conductivity.
The main idea is the existence of an {\it incompressible
quantum fluid} at specific rational values of the filling.
Repelling electrons assume the most ``symmetric''
configuration compatible with their density,
such that the ground state has a gap and such that
there are no phonons.
This picture clearly applies to the case of completely filled Landau
levels (integer Hall effect).
Compressions would excite electrons to higher levels, but these are
suppressed by the large cyclotron energy.
The Hall conduction is due to an overall rigid motion
of the droplet of quantum fluid, which yields \sig~ for integer $\nu$.
In this case, Coulomb repulsion can be neglected since its typical scale
is much smaller than the gap.
Moreover, topological arguments have been developed to show that
the free-electron picture is robust in the presence of impurities
\ref\wu{Y.-S. Wu, {\it ``Topological Aspects of the
Quantum Hall Effect''}, in {\it ``Physics, Geometry and Topology''},
H. C. Lee ed. , Proc. Banff Conf. 1989, Plenum Press, New York (1990).}.

At filling fractions $\nu=1/m$, with $m$ an odd integer, Laughlin's
wave function
\eqn\lauwf{\Psi({\bf x_i})= \prod_{i<j} {(z_i-z_j)^{m}}
\ {\rm e}^{\displaystyle -{1\over 2\ell^2}\sum_{i} |z_i|^2} ,}
($z_j=x_j + i y_j$ is the position of $j$-th electron and
$\ell=\sqrt{2 \hbar c / e B}$ is the magnetic length)
exhibits a similar incompressibility
\ref\laugha{R. B. Laughlin, {\it ``Elementary Theory: the
Incompressible Quantum Fluid''}, in \qhe .}.
It correctly approximates the numerical ground state for a
large class of repulsive interactions, and excitations
have a finite gap
\ref\halb{F. D. M. Haldane, {\it ``The Hierarchy of Fractional States
and Numerical Studies''}, in \qhe .}.
Laughlin developed the incompressibility picture, as well as the
properties of excitations, by interpreting $|\Psi|^2$
as a classical probability distribution for a two-dimensional
Coulomb gas of charges, the {\it plasma analogy} \laugha.
For $\nu=1/m$, this plasma is a liquid, the charges are completely
screened and stability amounts to local electrical neutrality.

A very interesting consequence of this theory is that the charged
excitations are {\it anyons}, {\it i.e.},
particles with fractional statistics $\theta/\pi=1/m$
\ref\wil{F. Wilczek ed., {\it ``Fractional Statistics and Anyon
Superconductivity''}, World Scientific, Singapore, (1990).},
and fractional charge $e/m$
\ref\laughb{R. B. Laughlin, ``Fractional Statistics in the Quantum
Hall effect'', in \wil .}.
Moreover, Haldane \hala~ and Halperin \halp~
constructed a whole hierarchy of new incompressible quantum
fluids for other filling fractions $\nu= {2\over 5}, {2 \over 7},
\dots$, by extending Laughlin's arguments on electrons
to anyons. The hierarchical structure of plateaus and
excitations has been observed experimentally and strongly supports
the entire picture.

As we already mentioned, we believe that this picture should be
supplemented with exact statements coming from symmetry.
We have in mind the exact fractional critical
exponents that follow from the infinite conformal symmetry of
two-dimensional critical phenomena \bpz. Actually, a formal analogy
between the quantum Hall effect and two-dimensional conformal
field theory was observed by Fubini
\ref\fub{S. Fubini, {\it Mod. Phys. Lett.} {\bf A6} (1991) 347.},
and Stone
\ref\sto{M. Stone, {\it Int. Jour. Mod. Phys.} {\bf B 5} (1991) 509.},
 and further developed in references
\ref\napo{S. Fubini and C. A. L\"utken, {\it Mod. Phys. Lett.} {\bf A 6}
(1991) 487; G. V. Dunne, A. Lerda and C. A. Trugenberger,
{\it Mod. Phys. Lett. } {\bf A 6} (1991) 2819;
C. Cristofano, G. Maiella, R. Musto and F. Nicodemi,
{\it Phys. Lett.} {\bf B 262} (1991) 88, {\it Mod. Phys. Lett.} {\bf
A 6} (1991) 1779, {\it Mod. Phys. Lett.} {\bf A 6} (1991) 2985.}.
This analogy is based on the similarity between the (holomorphic part
of the) Laughlin wave function \lauwf\ and the correlators of vertex
operators in a conformal theory with central charge $c=1$.

Further extensions of this analogy have produced new results
\ref\moore{G. Moore and N. Read, {\it Nucl. Phys.} {\bf B 360} (1991)
362; {\it ``Fractional Quantum Hall Effect and nonabelian statistics''},
YCTP-P5-92 preprint, to appear in proc. Sixth Yukawa Int. Symp.,
Kyoto (July 1991).}.
A new type of incompressible fluid was suggested at $\nu=1/2$,
which has been recently observed experimentally in double layer
samples
\ref\gww{M. Greiter, X. G. Wen and F. Wilczek, {\it Nucl. Phys.} {\bf
B 374} (1992) 567, and {\it ``Paired Hall States in Double Layer
Electron Systems''}, IASSNS-HEP-92/1 preprint.}.
Moreover, the possibility that the excitations in this
system possess non-abelian statistics is currently being debated \moore
\gww.

These interesting developments call for an explanation of this
analogy from a direct analysis of the microscopic physics
of the quantum Hall effect. At first sight, this looks
rather odd, given that conformal symmetry acts in $(1+1)$-dimensional
relativistic field theory, whereas the problem at hand is
$(2+1)$-dimensional and non-relativistic.
A first positive indication is that the Hilbert space of electrons
restricted to a given Landau level is the Bargmann space of
analytic functions
\ref\girjb{S. M. Girvin and T. Jach, {\it Phys. Rev.} {\bf B 29} (1983)
5617.}
\ref\itz{C. Itzykson, {\it ``Interacting electrons in a Magnetic Field''}
, in {\it ``Quantum Field Theory and Quantum Statistics,
Essays in Honor of 60th Birthday of E. S. Fradkin''},
A. Hilger, Bristol, (1986).}.
This projection produces a dimensional reduction of the particle
phase space from four to two dimensions.

In Section 2, we shall indeed show that there exists an
infinite-dimensional symmetry in the free electron theory of
Landau levels. However, the relevant algebra is not the
Virasoro algebra but a quantum version of the algebra of area-preserving
diffeomorphisms. These types of quantum algebras
are called $W_{\infty}$ and are currently the subject of active
investigations in two-dimensional gravity
\ref\wref{E. Witten, {\it ``Ground Ring of Two Dimensional String
Theory''}, IASSNS-HEP-91/51 preprint;
I. Bakas, {\it Phys. Lett.} {\bf B 228} (1989) 57;
C. N. Pope, X. Shen and L. J. Romans, {\it Nucl. Phys.}
{\bf B 339} (1990) 191.}.

In Section 3, the incompressibility of completely filled
Landau levels is related to the infinite symmetry. Namely,
the ground state is shown to be annihilated by infinitely many (in the
thermodynamical limit) symmetry generators. This is analogous
to the invariance of the vacuum in conformal field theory, {\it i.e.}
the highest-weight conditions.

In Section 4, the symmetry characterization of incompressibility
is extended to the Laughlin ground state \lauwf\ at
$\nu=1/m$. It is convenient and customary in this case to
limit the theory to the lowest Landau level, and to introduce
an effective short-range interaction for which the Laughlin
wave function is an exact incompressible eigenstate \hala \halb.
This interaction deforms the infinite symmetry algebra of the free case.
While the full properties of this theory, and of the deformed algebra,
are not discussed in this paper, we prove that the Laughlin
wave function for $\nu=1/m$ satisfies again the highest weight
conditions as in the $\nu= n$ case.

Finally, in the conclusions we discuss excitations and a geometrical
description of the fluid.
\vfill 
\eject 
\newsec{Infinite conserved charges in the non-interacting theory}
\bigskip
\noindent{\bf 2.1 Landau levels}
\bigskip
Let us begin by reviewing some basic facts about non-interacting
electrons in an external uniform magnetic field $B$ orthogonal
to the plane \girjb \itz.
We first discuss the one-body problem and shall deal later
with the many-body situation
and second quantization. The action $S$, and the Hamiltonian $H$,
of an electron of charge $e$ and mass $m$ are:
\eqn\elec{\eqalign{S &= \int dt \left ( {1\over 2} m {\bf v}^2 +
{e\over c} {\bf v{\cdot}A} \right ), \cr
H &= {1\over 2} m {\bf v}^2 = {1\over{2m}}
\left( {\bf p} - {e\over c}{\bf A} \right )^{2} \ , \cr}}
where we choose \footnote{*}{Our results
will naturally be gauge-independent. We shall comment later on
the formulas for other gauges.}
the symmetric gauge ${\bf A}={\displaystyle{B\over 2}}(-y,x)$.

At the semiclassical level (exact in this case),
the classical circular motion of
cyclotron frequency $\ \omega= e B/ mc\ $  gives rise to quantization
of the (kinetic) energy $E= \hbar \omega n$ and thus to quantized
radii $r_n$. The corresponding orbits enclose a flux multiple of the
quantum unit of flux $\Phi_{0} = h c /e$, {\it i.e.}
$\pi r_n^{2} B = n \Phi_{0}$.
The center of the orbit is instead free, thus excitations of arbitrary
angular momentum cost no energy and are degenerate.
The flux quantization implies that the magnetic field has an associated
unit of length, the {\it magnetic length} $\ell=\sqrt{2 \hbar c / e B}$.
{}From now on we choose units $\hbar=c=\ell=1\ $, so that $eB=2$ and
$\Phi_{0}=2\pi/e$. Moreover, for convenience we take $m=e=1$ in these
units, thus $\omega=2$.

At the quantum level, the canonical momentum is realized
as ${\bf p}= -i \nabla$ acting on functions of the
coordinate ${\bf x}=(x,y)$. The Hamiltonian and (canonical)
angular momentum can be written in terms of a pair of independent
harmonic oscillators:
\eqn\llh{\eqalign{H &= \ 2 a^{\dag }a + 1\ , \cr
J &= \ b^{\dag }b -a^{\dag }a \ ,\cr }}
In complex notation $z=x+iy$, $\bar z=x-iy$, $\partial =\partial
/\partial z$, $\bar \partial =\partial /\partial \bar z$,
these operators are
\eqn\eab{\eqalign{a &={z\over 2}+\bar \partial  \ ,
\qquad a^{\dag } ={\bar z \over 2}-\partial  \ , \cr
b &={\bar z \over 2}+\partial  \ ,
\qquad b^{\dag } ={z \over 2}-{\bar\partial} \ ,\cr }}
and satisfy the commutation relations
\eqn\eccab{[a, a^{\dag }] = 1\ ,\qquad
[b, b^{\dag }]=1 \ ,}
with all other commutators vanishing. Starting from the vacuum
$\Psi_{0}={1\over\sqrt{\pi}}\exp\left(-\vert z\vert^2/2\right)$,
satisfying $a \Psi_{0}= b \Psi_{0} = 0$, one indeed
finds energy $(a^{\dag})$-excitations, the Landau levels, which
are infinitely degenerate with respect to the angular momentum
$(b^{\dag})$-excitations.
The general wave function of energy $n$ and angular
momentum $l$ is given by
\eqn\epsi{\Psi_{n,l}({\bf x}) =
{{(b^{\dag})^{l+n}}\over{\sqrt{(l+n)!}}}
{{(a^{\dag})^{n}}\over{\sqrt{n !}}}
\Psi_{0} ({\bf x})=\ {\sqrt{n!\over{\pi (l+n)!}}}
\ z^{l} L_{n}^{l}( |z|^{2}) \ {\rm e}^{\scriptstyle -{|z|^2\over 2} }, }
where $L_{n}^{l}(x)$ are the generalized Laguerre polynomials and
$n \ge 0, l+n \ge 0 \ $.
For example, the wave functions of the lowest Landau level
satisfy $a \Psi_{0,l}=0$, and read
\eqn\psigs{\Psi_{0,l}({\bf x})= {z^{l} \over {\sqrt{ l !}}}
{1\over{\sqrt{\pi}}} \ {\rm e}^{\scriptstyle -{|z|^2\over 2} } .}
These functions are peaked at circular orbits, such that the
degenerate levels have an onion-like structure. The degeneracy
$N_{\cal A}$ of levels occupying an area $\cal A$ is equal to the flux
through it in quantum units,
$N_{\cal A}= \Phi/{\Phi_{0}}=B{\cal A}/2\pi$,
thus the degeneracy is finite in a finite sample.

The many-body problem of $N$ electrons will clearly be described in
first quantization
by taking $N$ copies of the previous operators labelled by an
index $i$: $a,b \to a_{i},b_{i}$.
In this case, the parameter $B$ acts as an external pressure,
because it controls the number of states and thus the density of
electrons per state. Actually, the latter is the correct quantum
measure of electron density, the {\it filling fraction} $\nu$
\eqn\nuin{\nu={\displaystyle{N\over{N_{\cal A}}}}\ \propto\ {1\over B}.
\qquad {\hbox{ (finite sample)}}}
In the infinite plane,
and within the lowest Landau level, the commonly used definition is:
\eqn\nup{\nu={\displaystyle {N(N-1)\over{2 J}}}. \qquad {\hbox{
 (infinite plane)}}}
Indeed, the area occupied by the electrons is proportional to the
total angular momentum $J= \sum_{i=1}^{N} J_{i}\ $, whose
minimal value is $N(N-1)/2$, by Fermi statistics.

To be more precise, the filling fraction $\nu$ has to be understood
as a thermodynamical expectation value, and in this sense
it appears in the conductivity, eq. \sig.
Therefore, eq. \nup~ indicates that $1/\nu$ is proportional
to the average value of $J$.
In order to control $J$ in the infinite plane with an external parameter,
we introduce a central harmonic potential which confines the particles
\eqn\confi{ H \rightarrow {\cal G}=\ H +{\lambda \over 2} \sum_{i=1}^N
\vert z_i\vert^2 \ ,}
where $\cal G$ is the Gibbs free energy.
Indeed, this modification is physical since it corresponds precisely
to the effect of the neutralizing positive charged background
\ref\trukiv{S. A. Trugman and S. Kivelson, {\it Phys. Rev.} {\bf B 31}
(1985) 5280.}.
The harmonic potential can be written in terms of $a$ and $b$'s and
therefore of $J$ and $H$
\eqn\confj{{\cal G}=(1+\alpha) H +\alpha J \ ,}
where $\lambda =\alpha (\alpha +2)$.
Thus, $\alpha$ is the external parameter conjugate to $H+J$,
{\it i.e.}, to $1/\nu$. Since $(J+H) \ge 0$
measures the area occupied by the electrons, $\alpha$ is
a confining pressure (infinite plane) and it is equivalent to
$1/B$ (finite sample).
Finally, note that the value of $\nu$ will be dominated by the angular
momentum of the ground state and fluctuations in $J$ of
$O(1,N)$ will describe the characteristic excitations.

In second quantization, the field operator is written as
\eqn\psisq{{\hat \Psi}({\bf x},t) = \sum_{n,k=0}^{\infty}
F_{k}^{(n)} \Psi_{n,k-n} ({\bf x})
\ {\rm e}^{\displaystyle - i(2n+1)t }\ ,}
in terms of the wave functions \epsi, the fermionic Fock
annihilators $F_{k}^{(n)}$, and creators $F^{(n)\dag}_{k}$, with
\eqn\ffdag{\{ F_{k}^{(n)}, F^{(m)\dag}_{l} \}= \delta_{n,m}
\delta_{k,l} \ .}
The $N$-body wave function is given by the Fock average
\eqn\psinbod{\Psi_{E}({\bf x_{1}}, \dots,{\bf x_{N}})=
\langle 0 |{\hat \Psi}({\bf x_{1}},0) \dots {\hat \Psi}({\bf x_{N}},0)
| N, E \rangle, }
where $|0\rangle$ is the Fock vacuum, and $|N,E\rangle$ the
corresponding Fock
state. Second quantization corresponds to the canonical quantization
of the non-relativistic Schr\"odinger field theory whose action is:
\eqn\ssft{ S = \int dt\ d^{2}{\bf x} \left ( i {\Psi}^{\dag}
{\displaystyle {\partial}\over{\partial t}} \Psi - {1\over{2}}
{\left ( {\bf D} \Psi \right )}^{\dag} \cdot{\left ( {\bf D}
\Psi \right )} \right ), }
where ${\bf D}= \nabla - i {\bf A}$. The commutation relations
\ffdag\ correspond to the canonical commutator
\eqn\psipsi{\{
\hat\Psi({\bf x},t),{\hat\Psi({\bf y},t)}^{\dag} \} = \delta^{(2)}
({\bf x} -{\bf y}) \quad ,}
and the conserved Hamiltonian $H$ and number operator $N$ are given
by
\eqn\handn{H= \int d^{2} {\bf x} \ {1\over{2}}
{\left ( {\bf D} \hat\Psi \right )}^{\dag} \cdot{\left ( {\bf D}
\hat\Psi \right )} , \qquad N= \int d^{2}{\bf x} \ {\hat\Psi}^{\dag}
\hat\Psi.}
\bigskip
\bigskip
\noindent
{\bf 2.2 Infinite quasi-local conserved charges}
\bigskip
We now return to discuss
the meaning of the $a$ and $b$ oscillators. The
operators $a$,$a^{\dag}$ are covariant derivatives,
as can be seen by restoring their $B$ dependence
\eqn\ddbar{
a={\bar\partial} + {B\over 4}z \ \ ,\ \ \
a^{\dag} =- \partial + {B\over 4}{\bar z} \ \ ,\ \ \
[a, a^{\dag} ] = {B\over 2}\ .}
The operators
$b$ and $b^{\dag}$ are generators of {\it magnetic translations},
{\it i.e.}, of translations combined with gauge transformations.
In a uniform magnetic field, there is clearly translation invariance.
Therefore, there should be two derivative operators commuting with
the Hamiltonian,
$\ [b,H]=[b^{\dag},H]=0\ .$
However, the choice of gauge apparently breaks translation
invariance. Therefore, a translation should be accompanied by
a compensating gauge transformation\footnote{*}{A general
discussion of symmetry in the presence of a background is given in
\ref\jacman{R. Jackiw and N. S. Manton, {\it Ann. Phys.}(NY) {\bf 127}
(1980) 257.}.}.
Actually, by exponentiation, the finite magnetic translations
$T_{\epsilon,{\bar{\epsilon}}}$
give rise to a projective representation of the translation group
\eqn\emagt{\eqalign{
T_{\epsilon,{\bar{\epsilon}}} \psi(z,{\bar z})\equiv &\
{\rm e}^{\displaystyle \epsilon b -{\bar\epsilon}{\bar b}} \Psi=
{\rm e}^{\scriptstyle{1\over 2}(\epsilon {\bar z}- {\bar{\epsilon}} z) }
\Psi(z+\epsilon,{\bar z}+{\bar{\epsilon}}), \cr
T_{\epsilon,{\bar{\epsilon}}} \ T_{\lambda,{\bar{\lambda}}}= &\
{\rm e}^{\scriptstyle{1\over 2}(\epsilon \bar\lambda -
\bar\epsilon \lambda )}\
T_{\epsilon +\lambda,{\bar\epsilon}+{\bar\lambda} } \ .\cr}}
The fact that the two correct translation operators commute with
$H$, but not among themselves $\ ([b,b^{\dag}]=1)\ $, accounts for the
infinite degeneracy of Landau levels on the unbounded plane.
It also gives other interesting effects in finite geometries
\napo.

It is worth stressing that the magnetic translation algebra is
gauge invariant, even if the explicit form of $b$ and $b^{\dag}$
in terms of $\partial_{i}$ and $x_{i}$ depends on the gauge.
As discussed in \fub
\ref\frad{E. Fradkin, {\it ``Field Theories of Condensed Matter
systems''}, Addison-Wesley, New York, (1991).}, in a general
gauge they are defined by
$\ b_{i}= \partial_{i} + i \Lambda_{i}\ $,
where $\Lambda_{i}({\bf x})$ are at most linear in the coordinates
and are determined by the conditions
$[b_i , p_j - A_j]=0, \quad i,j=1,2$. These imply
$\ [b_1 ,b_2 ]=[{p_1 - A_1},{p_2 -A_2}]= -iB,\ $
which establishes manifest gauge invariance.
Thus far, we discussed well-established matters.

We would now like
to make a new, yet simple, remark:
{\it there are more operators which commute with the
Hamiltonian.}
They are given by polynomials of the non-commuting operators $b$ and
$b^{\dag}$, defined by:
\eqn\defl{\eqalign{{{\cal{L}}_{n,m}} &\equiv (b^{\dag})^{n+1}\ b^{m+1},
\qquad n,m \geq -1, \cr
[{{\cal{L}}_{n,m}} &, H] = 0. \cr }}
These generate {\it quantum area-preserving
diffeomorphisms} (up to gauge transformations). Indeed, from the
definitions \eab\ and \eccab\  we deduce that
\eqn\em{ [{{\cal{L}}_{n,m}},{{\cal{L}}_{k,l}}]=\hbar \left( (m+1)(k+1)-
(n+1)(l+1) \right ) {{\cal{L}}_{n+k,m+l}} + O({\hbar}^{2}),}
where the missing terms correspond to contractions of more derivatives,
and hence
have higher powers of $\hbar$. The $O(\hbar)$ term, the classical
algebra, identifies our algebra as that of area-preserving
diffeomorphisms or isometries. This is called $w_{\infty}$ in the
context of two-dimensional gravity \wref. Quantum deformations of
this algebra are called, collectively, $W_{\infty}$-algebras. There
are many such deformations, which differ in the higher order terms in
$\hbar$ \footnote{*}{They can be changed by
modifying the ordering of $b$, $b^{\dag}$ in the definition \defl.}.

The ${\cal{L}}_{n,m}$ generically involve powers of derivatives
higher than one, and therefore are not generators of local
coordinate transformations on the wave function; they are
{\it quasi-local} operators. The generating function of ${\cal{L}}_{n,m}$
is actually the finite magnetic translation discussed before \emagt.
The only local coordinate transformations are the translations
${\cal{L}}_{0,-1}, {\cal{L}}_{-1,0}$,
and rotations $({\cal{L}}_{0,0}-a^{\dag}a)$
with which we started.

For $N$ particles, the first-quantized
form of the generators ${\cal{L}}_{n,m}$ is
\eqn\defln{{\cal L}_{n,m} \equiv
\sum_{i=1}^{N}\  (b_{i}^{\dag})^{n+1} b_{i}^{m+1}, \qquad n,m \geq -1.}
Note that in the thermodynamical limit they become an infinite set of
independent charges.

The second-quantized formulae might help to convince the reader that
the ${\cal{L}}_{n,m}$ are bona-fide conserved charges arising from
conserved (quasi)-local currents. Their second quantized expression is
\eqn\lsecq{{\cal{L}}_{n,m}=\int d^{2} {\bf x} \ {\hat{\Psi}}^{\dag}
\ (b^{\dag})^{n+1} (b)^{m+1} \ {\hat{\Psi}},}
where $\hat{\Psi}$ are the field operators introduced in eq. \psisq.
Since \hbox{$ [(b^{\dag})^{n+1}(b)^{m+1},H]=0$},
the ${\cal{L}}_{n,m}$ are conserved charges. This can be cast
in the usual form of current conservation
\eqn\currc{\partial_{0} J_{(n,m)}^{0} + \partial_{i} J_{(n,m)}^{i}=0,}
where
\eqn\currd{\eqalign{
{J^{0}_{(n,m)}}({\bf x}) &\equiv \ {\hat{\Psi}}^{\dag}
(b^{\dag})^{n+1}(b)^{m+1} {\hat{\Psi}},  \cr
{J^{j}_{(n,m)}}({\bf x}) &\equiv  {i\over2} \left [ {\left( D_j
{\hat{\Psi}} \right )}^{\dag} (b^{\dag})^{n+1} b^{m+1} {\hat{\Psi}}
-{\hat{\Psi}}^{\dag} (b^{\dag})^{n+1} b^{m+1} D_j {\hat{\Psi}}
\right ], \cr}}
Current conservation follows from the operator equation of motion
of the Hamiltonian \handn, and the mentioned fact that
$b$, $b^{\dag}$ commute with the covariant derivatives.
It is worthwhile to point out that strictly speaking, the
observables have to be real, and they correspond to the eigenvalues
of $Re({\cal{L}}_{n,m})$ and $Im({\cal{L}}_{n,m})$.
\bigskip
\bigskip
\noindent{\bf 2.3 Classical symmetry}
\bigskip
In order to gain some geometrical intuition,
let us clarify the classical origin of the symmetry in our problem.
In relativistic field theory we usually look for symmetries as those
coordinate transformations that leave the action invariant.
However, from a Hamiltonian point of view, symmetries are
canonical transformations that leave the Hamiltonian invariant.
Besides coordinate transformations, there are more general
transformations which change momenta independently of the coordinates.
These are usually disregarded in field theory, since they act non-locally
in the Hilbert space, and therefore are difficult to implement at the
quantum level\footnote{*}{
Nevertheless, we quote the example of the B\"acklund transformation
in the canonical quantization of Liouville theory
\ref\gand{G. I. Ghandour, {\it Phys. Rev.} {\bf D 35} (1987) 1289.}.}.
As we now show, the $W_\infty$-symmetry in our problem is the quantum
version of a canonical symmetry of the classical Hamiltonian \elec.

To begin, we briefly recall some basic facts about
canonical transformations in the simplest example of a two-dimensional
phase space $(q,p)$.
Canonical transformations are diffeomorphisms
of the phase space which preserve the symplectic two-form
$\omega= dq \wedge dp$, {\it i.e.} the area form
\ref\gold{H. Goldstein, {\it ``Classical Mechanics''}, Addison-Wesley,
Reading, 1980.}.
The infinitesimal transformations have a generating function
\hbox{$F=F(p,q)$},
\eqn\canon{\eqalign{ Q &=q+ \delta q , \qquad
\delta q= \{ q, F \}_{PB} ={\partial F \over{\partial p}}, \cr
P &=p+ \delta p , \qquad
\delta p= \{ p, F \}_{PB} =-{\partial F \over{\partial q}}, \cr}}
where $\{$ , $\}_{PB}$ are the Poisson brackets, $\{q, p\} =1$.
The ``point'' transformations, $\delta q=f(q)$,  are given by
$ F=p f(q)\ $, under which $p$ transforms as a derivative.
A complete basis of generators is given by the full power series
expansion of $F(p,q)\ \ , \ F_{n,m}= -q^{n+1} p^{m+1} $.
The Poisson brackets of two such generators
give a representation of the classical $w_\infty$ algebra \wref.

Let us now explain how this algebra can arise in our problem,
which has a {\it four-dimensional \/} phase space
$\{ x,p_{x},y,p_{y} \}$.
The classical Hamiltonian \elec~ can be written as
\eqn\calssh{H = {1\over 2} \left ( (p_{x} + y)^{2} +
(p_y -x)^{2} \right ) = {1\over 2} \left ( v_{x}^{2} + v_{y}^{2}
\right ),}
The most general canonical transformations which leave it invariant
are generated by
\eqn\deltah{\eqalign{{\cal L}^{(cl)}_{n,m}\
&{\bf ( x, p)} = (b^{\dag})^{n+1} b^{m+1}\ \ ,\ \ \ \
b = {1\over 2} \left ( (p_{y}+x) + i (p_{x} - y) \right ), \cr
\delta H &= \{ H,{\cal{L}}^{(cl)}_{n,m} \}_{PB} =0 \ ,\cr} }
as is clear from the classical limit of the discussion in section 2.2,
${\cal L}_{n,m} \rightarrow {\cal L}^{(cl)}_{n,m}\ $, and
$[{\hbox{  ,  }}] \to i\hbar \{ {\hbox{  ,  }} \}_{PB} + O({\hbar}^2)$.
This also implies a crucial property of these transformations,
namely, that they leave invariant the two combinations
$v_x= {\rm Im}\  a, \ v_y=-{\rm Re}\  a\ $, where \hfill\break
$a = {1\over 2} \left ( -(p_{y}-x) + i (p_{x} + y) \right )$,
{\it i.e.},
\eqn\clastr{\delta (p_{x}+y) = 0,\qquad
\delta (p_{y}-x)=0 \ .}
Therefore, the ${\cal L}^{(cl)}_{n,m}$ act non-trivially only on
a two-dimensional subspace of the four-dimensional phase space,
characterized by constant energy. It is a peculiar property of
this problem that a two-dimensional subspace admits a symplectic
structure in terms of $b$ and $b^{\dag}$.
Thus, the infinite canonical transformations on this subspace
are infinite symmetries and satisfy the $w_{\infty}$
algebra
\eqn\emcl{ \{ {\cal L}^{(cl)}_{n,m},{\cal L}^{(cl)}_{k,l} \}_{PB}=
\left( (m+1)(k+1)- (n+1)(l+1) \right ) \ {\cal L}^{(cl)}_{n+k,m+l} \ ,}
which is the classical limit of \em.

Upon quantization, something peculiar happens: within each Landau level,
the previous two-dimensional subspace is identified with the
actual coordinate space of the electrons.
As emphasized in
\ref\topcs{G. V. Dunne, R. Jackiw and C. A. Trugenberger, {\it Phys.
Rev.} {\bf D 41} (1990) 661; G. V. Dunne and R. Jackiw, unpublished
(April 1992).},
this {\it phase-space reduction \/} can be obtained by taking the
limit $m\rightarrow 0$ of the action \elec:
\eqn\srest{ {\tilde S}= \lim_{m\to 0} S =\
\oint_{orbit} dt\ {\dot{\bf x}}\cdot {\bf A} = \Phi =
\int dt\ \left( {\dot y} x - {\dot x} y \right) \ , }
where $\Phi$ denotes the flux piercing the area encircled by the orbit.
The action $\tilde S$ (``Chern-Simons mechanics'') describes classically
the residual degree of freedom within each Landau level
(the constant energy of the level is renormalized to zero).
Note that $\tilde S$ in \srest\ is of first order in time derivatives,
thus the symplectic structure is immediately evident and implies
that only one of the original coordinates remains a coordinate in
the Hamiltonian sense - the other becomes the conjugate momentum.
This is the advertised phase-space reduction induced by the
external magnetic field.

Since $\tilde H=0 \ $, the symmetries of the Hamiltonian are
all canonical transformations of the reduced phase space.
Equivalently, the symmetries of the action $\tilde S$
are the area preserving diffeomorphisms of the coordinate space, which
leave $\Phi$ invariant.
\bigskip
\bigskip
\noindent
{\bf 2.4 The full $W_{\infty}$ algebra}
\bigskip
We now present some additional properties of the full quantum algebra
and discuss its relation to the standard nomenclature
\wref.
The full commutator \em\ follows from the Leibnitz differentiation
rule ($\hbar=1$)
\eqn\comutls{[\ {\cal{L}}_{n,m}\  ,\  {\cal{L}}_{k,l}\  ] =
\sum_{s=0} ^{Min(m,k)} { (m+1)! (k+1)! \over{(m-s)! (k-s)! (s+1)!}} \
{\cal{L}}_{n+k-s,\ m+l-s} -
(m \leftrightarrow l ,n \leftrightarrow k)\ }
Special cases are given by:
\eqn\subal{\eqalign{
{\it i)} \qquad\qquad\qquad\qquad &
[\ {\cal{L}}_{n,0} , {\cal{L}}_{k,0}\ ] = (k-n){\cal{L}}_{n+k,0}\ ;\cr
{\it ii)} \qquad\qquad\qquad\qquad &
[\ {\cal{L}}_{0,n} , {\cal{L}}_{0,k}\  ] = (n-k){\cal{L}}_{0,n+m}\ ;\cr
{\it iii)}\qquad\qquad\qquad\qquad &
[\ {\cal{L}}_{n,n} , {\cal{L}}_{k,k}\ ] = 0 \qquad\quad\quad \quad
   {\rm ( Cartan\  subalgebra)}\ ;\cr
{\it iv)} \qquad \qquad \qquad \qquad &
[\ {\cal{L}}_{0,0} , {\cal{L}}_{n,m}\  ] = (n-m){\cal{L}}_{n,m}\ .
\cr} }
The first two are Virasoro algebras, but the negative $(n < -1)$
modes are missing. Later, we shall show that
$\ {\cal{L}}_{n,0} = \sum_{i} z_{i}^{n+1} \partial_{i}\ $,
when acting on analytic functions of the lowest level \psigs:
these look like the conformal transformations which were assumed in
references \napo; however, there are important differences.
The missing negative modes correspond to singular transformations
at the origin, which are not allowed on our two-dimensional
Hilbert space ($(b^{\dag})^{-k}$ is not defined). On the
other hand, Virasoro transformations are reparametrizations
of the circle and can be extended to singular transformations
of the plane. Moreover, the incompleteness of the algebra
prevents us from considering possible central extensions. In the
Virasoro algebra $L_{k}^{\dag} = L_{-k},\ $ but in our case
$\ {\cal{L}}_{n,0}^{\dag} = {\cal{L}}_{0,n}$, and these do not
satisfy a Virasoro algebra with the ${\cal{L}}_{k,0}$.

The standard notation of $W_{\infty}$ algebra stresses its relation
to the Virasoro algebra. Indeed, we can express our operators as
\eqn\weqv{V_{n}^{i} = -\ {\cal{L}}_{n+i,i}, \qquad i \ge -1, \quad
n \ge -i-1 \quad ,}
where
the index $i+2$ corresponds to the conformal spin. For $i=0$, the
Virasoro sub-algebra is satisfied by the Fourier modes of the spin-two
stress-tensor. For $i=1$, one has the Zamolodchikov spin-three
current, and similarly for higher spins. The index $n$ is the
conformal dimension, namely the eigenvalue of ${\cal{L}}_{0,0}$.
In our case, the Fourier modes $\ V_{n}^{i}\ $
are bounded from below $(n \ge -i-1)$, while in conformal field
theory they are unbounded
$\ \{ V_{n}^{i} , \quad n \in {\bf Z}, \quad i \ge 0
\quad {\hbox{ in CFT}} \}\ .$
Therefore, our algebra corresponds to the so-called ``wedge''
$W_{\Lambda}= \{ V_{n}^{i} , |n| \le i+1 \}$, plus the positive
modes $n > i+1$. In the standard notation of $W_{\infty}$, it
reads (restoring $\hbar$ for comparison with reference \wref ):
\eqn\comuvs{[V_{n}^{i},V_{m}^{j}] = \hbar\lambda V_{n+m}^{i+j}
+ \hbar^2\eta V_{n+m}^{i+j-1} + \hbar^3\gamma
V_{n+m}^{i+j-2} + \dots + \hbar^{i+j+1-q}\ \rho V_{n+m}^{q} \quad ,}
where $q=Min(Max(i-m,j), Max(i,j-n))$. In eq. \em,  we omitted
the structure constants, which differ from those of reference \wref,
apart from the classical one. As anticipated, their form
depends on the choice of quantum ordering and it is largely
arbitrary, provided the Jacobi identity is satisfied. Therefore,
the notion of $W_{\infty}$ really stands for a large class of
quantum algebras.
As in the case of the Virasoro subalgebra, $W_{\infty}$ admits in
general a central term $\delta^{ij}\delta_{n+m,0} c_{i}(n)$.
In its minimal form \wref, $c_{i}(n)$
vanish inside the wedge, thus it cannot be added to our
algebra.
\vfill 
\eject 
\newsec{Infinite symmetry and incompressibility of the ground
state at integer fillings}
\bigskip
In the previous section, we revealed an infinite-dimensional
algebra in the non-interacting theory; we now investigate its
relevance for the integer Hall effect.
Up to now, we have been concerned primarily with the commutators
of the generators with the Hamiltonian. Before we can assert
that there is a symmetry in the quantum theory, we should
discuss the action of the generators on the states, in particular
the ground state $|\ \Omega\ \rangle$.
We shall find an implementation of the symmetry analogous
to conformal field theory. The ground state satisfies the
highest weight conditions, {\it i.e.}, it is annihilated
by an infinite subset of ``lowering'' operators, while the
others create excitations. In conformal field theory, this
condition is
\eqn\hwc{L_{n} |\ \Omega\ \rangle=0, \qquad n \ge -1 \quad .}

Let us first understand the action of the ${\cal{L}}_{n,m}$
in Fock space.
Their second-quantized expression is
obtained by inserting the
field operator \psisq\ into definition \lsecq
\eqn\lnm{\eqalign{{\cal{L}}_{n,m} &= \sum_{k,j=0}^{\infty}
\ \sum_{l,i=0}^{\infty} F^{(k)\dag}_{j}
F_{i}^{(l)} \int d^{2} {\bf x}\ {\overline{\Psi_{0}}({\bf x})}
{{a^{k}}\over{\sqrt{k!}}}
{{b^{j}}\over{\sqrt{j!}}} (b^{\dag})^{n+1} b^{m+1}
{{{(a^{\dag})}^{l}}\over{\sqrt{l!}}}
{{(b^{\dag})^{i}}\over{\sqrt{i!}}} \Psi_{0}({\bf x}) \cr
&=\ \sum_{r=0}^{\infty} {\cal{L}}_{n,m}^{(r)}, \cr
{\cal{L}}_{n,m}^{(r)} &= \sum_{k \ge m+1}
{\sqrt{k! (k+n-m)!}\over{(k-m-1)!}}\ F^{(r)\dag}_{k+n-m}
F_{k}^{(r)}. \cr}}
The ${\cal{L}}_{n,m}$ split into identical copies
${\cal{L}}_{n,m}^{(r)}$ acting independently within each level
$(r)$. The bilinears $F_{k+s}^{\dag}F_{k}$ show that
${\cal{L}}_{n+s,n}$, for fixed $s$, displaces electrons by
increasing $(s > 0)$ or lowering $(s < 0)$ their angular
momentum. They act as {\it raising} and {\it lowering} operators
within each level. Clearly, the bilinears
$F_{k+s}^{\dag}F_{k}$ give the simplest bosonic description
of the Hilbert space at each level, because any fermionic state
can be connected to any other by their repeated action.
Therefore they achieve a ``bosonization'' of the non-relativistic
fermion theory in each level. Moreover, the ${\cal{L}}_{n,m}^{(r)}$
similarly span the single-level Hilbert space, given that
$\{ F^{(r)\dag}_{k+s} F_{k}^{(r)} \}$
and $\{ {\cal{L}}_{n,m}^{(r)} \}$ are
equivalent sets. Actually, eq. \defln\ can be inverted
within the $r$-th level as follows:
\eqn\boso{F^{(r)\dag}_{a} F_{b}^{(r)} = {1\over{\sqrt{a! b!}}}
\sum_{k=-1}^{\infty} {(-1)^{k+1} \over{(k+1)!}}\ {\cal{L}}_{a+k,
b+k}^{(r)} \quad .}
As a consequence, the algebra of the ${\cal{L}}_{n,m}$ is a
``spectrum-generating algebra'' for the angular momentum.

Next, we discuss some physical issues of the integer Hall effect.
The non-interacting electron theory considered so far is
generically unphysical. The Coulomb interaction lifts the
degeneracy of the Landau levels such that the properties of
the ground state depend on the value of the filling fraction
$\nu$. However, for integer filling $\nu=1,2,\dots$, the
non-interacting theory makes physical sense \footnote{*}{A more
precise statement based on the renormalization group will be
made in section~4.} \wu.
Under the
experimental conditions of large magnetic fields, the energy
gap between Landau levels is larger than the typical
energy scale of the Coulomb interaction:
\eqn\coul{\displaystyle{ {\hbar}{eB\over{mc}} > {e^{2}
\over{\kappa \ell}},}}
where $\kappa$ is the dielectric constant.
Therefore, for integer $\nu$, the Landau levels retain their character,
the ground state is unique and has a gap: this
corresponds to an {\it incompressible} quantum fluid \laugha.
Transitions of one or more electrons to higher levels would
reduce the angular momentum, {\it i.e.}, would compress the
fluid; however, these are forbidden by the large gap.

Moreover, angular momentum excitations, {\it i.e.},
decompressions, are controlled in our geometry by the
confining potential \confj\ discussed in section 2.1. Their
energy is given by the Gibbs energy ${\cal{G}}=\alpha J=
\alpha {\cal{L}}_{0,0}$ (within each Landau level), where
the value of $\alpha \propto 1/B$ allows for completely
filled Landau levels, say the first one. Angular
momentum excitations become energy excitations,
and we achieve a very close analogy with conformal field theory,
where similarly $H=L_{0}$. Indeed, the incompressible ground state
at $\nu=1$ satisfies highest weight conditions analogous to
\hwc, as we now show.

The ground state is given by
\eqn\gssq{|\ \Omega\ \rangle\ =\ |N,\nu=1\rangle =F^{(1)\dag}_{0}
F^{(1)\dag}_{1} \dots
F^{(1)\dag}_{N-1} |\ 0\ \rangle }
in second quantization, and by
\eqn\gsfq{\Psi_{\nu=1}({\bf x_{1}},\dots,{\bf x_{N}})= \prod_{i
<j}^{N} (z_i - z_j )\ e^{- {1\over 2}{\displaystyle{ \sum_{i=1}^{N}
|z_{i}|^{2}}}}}
in first quantization. The highest weight conditions read
\eqn\invac{{\cal{L}}_{n,m} \Psi_{\nu=1} =0,
\qquad {\hbox{ for}} \quad
-1 \le n < m, \quad m \ge 0 \quad,}
and follow from the nature of ${\cal{L}}_{n,m}$.
Indeed, ${\cal{L}}_{n,m} \to
{\cal{L}}_{n,m}^{(1)}$ when acting on the lowest level, where
they lower the angular momentum for $n < m$ and therefore
compress the fluid. Thus, they vanish on a completely filled
state. Moreover, they cannot generate compression transitions to
higher levels, energetically forbidden, because they commute
with the Hamiltonian.

Eqs. \invac\ precisely state
{\it the incompressibility conditions of the filled Landau
level}. Geometrically, they express the symmetry, {\it i.e.},
stability of this fluid \footnote{*}{Note that
the condition of magnetic translation invariance,
${\cal{L}}_{-1,0} \Psi = 0$, has
been previously recognized as a necessary condition for
incompressibility of the ground state \trukiv \itz.}.
Similarly, in conformal field theory, the highest weight
conditions \hwc\ express the invariance of the vacuum under
conformal transformations regular at the origin.

More precisely, let us compare both conditions
in the customary $W_{\infty}$ notation (see Section 2.4):
\eqn\whwc{V_{n}^{i} |\ \Omega\ \rangle = 0, \qquad i=0,1,\dots,
\ -i-1\le n <0 \quad . \quad {\hbox{ (Hall effect)}}}
These conditions have to be compared with those corresponding to the
$W_{\infty}$-symmetric vacuum in conformal theory:
\eqn\whwcc{V_{n}^{i} |\ \Omega\ \rangle = 0, \qquad i=0,1,\dots,
\ -i-1 \le n < \infty \quad . \quad {\hbox{ (CFT)}}}
Note that all conditions in \whwc\ are contained in
\whwcc. In the Hall case, one finds by inspection that there are
$O(N^{2})$ non-trivial conditions for $N$ electrons, such
that the ground state is infinitely symmetric in the thermodynamical
limit.

The same incompressibility conditions \invac\ are easily
extended to $k$-completely filled Landau levels. They read:
\eqn\invacl{{\cal{L}}_{n,m} |N,\nu=k\rangle =0, \qquad {
\hbox{ for}} \quad -1 \le n < m, \quad m \ge 0,\quad k=1,2,\dots \quad .}
Indeed, using second quantization, for $N=kN_{L}$ electrons:
\eqn\sqeq{|N, \nu =k \rangle= \prod_{r=1}^{k} \left(
F^{(r)\dag}_{0} F^{(r)\dag}_{1} \dots
F^{(r)\dag}_{N_{L}-1} \right) | 0 \rangle, }
thus,
\eqn\vacinv{{\cal{L}}_{n,m} |N,\nu= k\rangle= \sum_{r=1}^{k}
(\dots)\ {\cal{L}}_{n,m}^{(r)}\ \left(
F^{(r)\dag}_{0} F^{(r)\dag}_{1} \dots
{F^{(r)\dag}_{N_{L}-1}} \right) | 0\rangle =0, }
by commuting operators for different levels.

In summary, we have shown that in the case of the integer
Hall effect, the incompressibility of the ground state corresponds
to an infinite symmetry under the $W_{\infty}$ algebra.
\vfill
\eject
\newsec{Infinite symmetry and incompressibility at fractional
fillings}
\bigskip
In this section, we show that the incompressibility of the
Laughlin wave function at
fractional fillings $\nu=1/m$, $m$ odd,
is again characterized by a set of highest weight conditions,
similar to the integer case.
\bigskip
\noindent
{\bf 4.1 Projection onto the lowest Landau level }
\bigskip
For fractional fillings, non-interacting electrons have access
to a large reservoir of degenerate states. One has to understand
how the repulsive interaction manages to arrange the electrons
in a collective ``symmetric'' state, such that the ground state
is unique and has a gap in the thermodynamical limit.
We already remarked that under experimental conditions,
the Landau level
gap is larger than the typical scale of the Coulomb
interaction, eq. \coul. A better estimate of the interaction
scale is given by the
size of the gap for the low-lying excitations above the $\nu=1/m$
ground state, which numerical simulations and
Laughlin's theory predict to be a few
percent of the Landau level gap \laugha \halb.
Therefore,
it is probably sufficient to limit the theory
to the lowest level. The Laughlin wave functions \lauwf\
belong indeed to this level, {\it i.e.}, they are of the general
form
\eqn\gpsilll{\Psi({\bf x_{1}},\dots,{\bf x_{N}})=
e^{\displaystyle{ -{1 \over 2} \sum_{i} |z_{i}|^{2}}}
\varphi (z_{1},\dots,z_{N}),}
where $\varphi$ is an entire analytic function of $z_{1},\dots,
z_{N}$ (see section 2.1). The projection onto the lowest level is
largely accepted in the literature for $\nu=1/m$, while it is
being debated in the case of hierarchical constructions at
fillings $\nu = p/q$
\ref\jain{J. K. Jain, {\it Phys. Rev.} {\bf B 41} (1990) 7653.}.

The projection leads us to the Bargmann space of coherent states
for the $b$, $b^{\dag}$ oscillators. Let us recall some of its
basic features \girjb \itz. The exponential factor in \gpsilll\
is absorbed into the measure and operators act only on the analytic
part of the wave function $\varphi (z)$
\eqn\barga{\Psi=\ e^{-{1 \over 2}\displaystyle{ z{\bar z}}}\ \varphi
(z), \qquad \langle
\phi\ |\ \varphi\rangle = \int {{dz d{\bar z}}\over{2\pi i}}
e^{\displaystyle{- z{\bar z}}} {\overline{\phi(z)}} \varphi (z).}
{}From eqs. \eab, $b$ and $b^{\dag}$ are indeed represented by
$z$ and $\partial$,
respectively. The orthogonality and completeness conditions for
coherent states are (see section 2.1)
\eqn\cohst{ |\ z\rangle= e^{\displaystyle{ {\bar z} b^{\dag}}} |\
\Psi_{0}\rangle,
\qquad \varphi(z)\ =\ \langle z\ |\ \varphi\rangle,}
\eqn\ideker{\langle z\ |\ \eta\rangle =
e^{\displaystyle{z {\bar{\eta}}}}\ ,\qquad
{\hbox{ (identity kernel)}}}
\eqn\comple{{\bf 1}\ =\ \int {{dz d{\bar z}}\over{2\pi i}}
e^{\displaystyle{-z {\bar z}}}\ |\ z\rangle\ \langle z\ |.}
Notice that ${\bar z}$ acts on the left and is the adjoint of
$\partial$, because one finds by partial integration that
$\langle \phi |{\partial} \varphi \rangle\
=\ \langle z\phi\ |\varphi\rangle$.
This confirms that the two-dimensional plane $(z,{\bar z})$
is a phase space for the one-level theory, in agreement with the
classical analysis of section 2.3.
Any operator $A$ is represented by an integral kernel $A(z,{\bar \eta})$
and acts as follows
\eqn\opker{(A\varphi)(z)\ =\ \langle z|A\varphi\rangle \ =\ \int
{{d{\eta}d{\bar{\eta}}}\over{2\pi i}} e^{\displaystyle{-\eta{\bar \eta}}}
A(z,{\bar \eta})\ \varphi(\eta).}
In particular, the kernel
$A(z,{\bar \eta})$ of a (quasi)-local operator $A$ is the same operator
applied to the identity kernel
\eqn\locker{A\ \varphi(z)\ =\ z^{k} {\partial}^{r} \varphi(z) \quad
\leftrightarrow \quad A(z,{\bar \eta})\ =\ z^{k}
{\left({\partial \over{\partial z}}\right)}^{r}
e^{\displaystyle{z {\bar \eta}}}
=\ z^{k} {\bar \eta}^{r} e^{\displaystyle{z {\bar \eta}}}. }
The projector $P_{0}$ onto the lowest level acts
on a generic wave function $\Psi$, setting
to zero its higher level components.
This is achieved by using the identity kernel
\eqn\proja{\Psi(z,{\bar z})\ =\ e^{{-{1\over2}}
\displaystyle{z{\bar z}}}
\Phi(z,{\bar z}),\qquad \qquad
\left( P_{0} \Phi \right) (z)\ =\ \int
{{d{\eta} d{\bar{\eta}}}\over{2\pi i}}
e^{\displaystyle{-\eta{\bar \eta}}}\
e^{\displaystyle{z{\bar \eta}}}\ \Phi(\eta,{\bar{\eta}}). }
Indeed, if
$\Phi=(a^{\dag})\ \phi(\eta,{\bar{\eta}})\ =\ \left(
{\bar{\eta}} - {\partial_{\eta}}  \right) \phi$,
it vanishes inside the integral by partial integration.
Similarly, the projection of local operators $V({\bar z},z)$
is obtained by moving all $\bar z$ powers to the left of
the $z$'s (normal ordering) and by replacing
${\bar z}\ \to \ \partial$ \girjb\ :
\eqn\projv{P_{0}\ V({\bar z},z)\ P_{0}\ =\ :V({\partial}
,z):\quad.}
For example,
\ $P_{0}\ |z|^{2}\ P_{0}\ =\ {\partial} z\ =\ 1+z\partial$.
The same result is obtained by applying $P_{0}$ in the
integral form \proja\ on both sides \itz.

The generators of area-preserving diffeomorphisms \defln\ take the
form
\eqn\lnmzzb{{\cal{L}}_{n,m}\ =\ \sum_{i=1}^{N} z_{i}^{n+1}
{\left(\partial\over{\partial z_{i}}\right)}^{m+1}\quad .}
As anticipated in section 2.4, the ${\cal{L}}_{n,0}$ actually generates
conformal transformations on analytic wave functions. Note that, as a
result of
the projection, the subset of local transformations
of our algebra is enlarged to include this infinite subset.
\bigskip
\bigskip
\noindent
{\bf 4.2 Effective interaction for $\nu=1/ m$}
\bigskip
A convenient parametrization of projected two-body interactions
is given by a ``partial-wave'' analysis or, equivalently, by a
short-range expansion
\footnote{*}{This is also reminiscent of the
operator product expansion in conformal field theory~
\ref\ag{L. Alvarez-Gaume, G. Sierra and C. Gomez, {\it ``Topics in
Conformal
Field Theory''}, in {\it ``Physics and Mathematics of Strings''},
L. Brink,
D. Friedan and A. M. Polyakov eds, World Scientific, Singapore (1990).}.}
\girjb \itz.
The previous rule of projection \proja\ tells us that the kernel of
a projected two-body operator, $V(|z_1- z_2|)$, is given by
\eqn\projen{\eqalign{\left(P_{0}\ V\ P_{0}\right)
(z_1 ,z_2 ; {\bar{\eta_1}}, {\bar{\eta_2}})\ &=\
\int d^{2}{\bf{\rho}}\ d^{2}{\bf{\xi}}\ e^{\displaystyle{
-\rho{\bar{\rho}}- \xi{\bar{\xi}}}}\ e^{\displaystyle{
z_1 {\bar{\rho}}+ z_2 {\bar{\xi}}}}\ V(|\rho -\ \xi|)
\ e^{\displaystyle{\rho{\bar{\eta_1}} +\ \xi{\bar{\eta_2}}}} \cr
&=\ e^{\displaystyle{{z_1+z_2\over{\sqrt{2}}}}}
\ e^{\displaystyle{ {{\bar{\eta_1}}+
{\bar{\eta_2}}\over{\sqrt{2}}}}} \sum_{n=0}^{\infty}
{e_{n} \over{n!}}\ {\left({z_1-z_2 \over{\sqrt{2}}} \right)}^{n}
{\left({{\bar{\eta_1}}-{\bar{\eta_2}} \over{\sqrt{2}}} \right)}^{n} ,
\cr}}
where the amplitudes $e_n$ of the partial waves are
\eqn\defen{e_{n}= {1\over{n!}} \int_{0}^{\infty} dr^{2}
e^{\displaystyle{-r^{2}}} r^{2n} V({\sqrt 2} r).}
The terms in the expansion of the kernel can be rewritten as
\eqn\srange{P_{0} V(1,2) P_{0}\ =\ \sum_{n=0}^{\infty} e_{n} V_{n}(1,2)
,\qquad V_{n}=\ |J_{12}=n\ \rangle\ \langle\ J_{12}=n\ |.}
Here, $V_{n}$ is the projection on the relative angular momentum
$J_{12}=n$. Indeed, when acting on a two-particle wave
function, it gives
\eqn\phipw{\varphi(z_1,z_2)\ =\ \sum_{n,m} C_{n,m}
{\left({z_1-z_2 \over{\sqrt{2}}} \right)}^{n}
{\left({z_1+z_2 \over{\sqrt{2}}} \right)}^{m}\ ,}
\eqn\vphipw{\left( V_{n}\varphi \right)(z_1,z_2)\ =\
{\left({z_1-z_2 \over{\sqrt{2}}} \right)}^{n}\ \sum_{m} C_{n,m}
{\left({z_1+z_2 \over{\sqrt{2}}} \right)}^{m}\ ,}
{\it i.e.}, it selects the term of $O\left( {(z_1 - z_2)}^{n} \right)$
in the expansion of $\varphi(z_1,z_2)$ around $\displaystyle{{z_1 +
z_2}\over 2}$. By antisymmetry, even powers never appear; thus, we
should only consider $V_{2k+1}$, $k=0,1,2,\dots$. An equivalent
notation \trukiv, valid only within expectation values, is
\eqn\deltpo{V_{n}(1,2) \sim {\Delta}^{n}\ \delta^{(2)}
({\bf x_1} -{\bf x_2}),\qquad <\Psi|V_{n}|\Psi>\ =\ \int d^{2}
{\bf x}\ {\overline \Psi({\bf x})}\  {\Delta}^{n} \Psi({\bf x}),}
where $\Delta$ is the Laplacian. For example,
the partial waves of the Coulomb interaction are
\eqn\coulv{H^{(c)}(1,2)\ =\ {1 \over{ |{\bf x_1}-{\bf x_2}|}}
\qquad \leftrightarrow \qquad e_{n}^{(c)}= {\Gamma(n+{1\over 2})
\over{n!\sqrt{2}}} \ {\buildrel n \to \infty \over \longrightarrow}\
{1 \over {\sqrt{2n}}}\quad .}
Note that
the $e_{n}^{(c)}$ are positive and monotonically decreasing.
Their overall dimensionful factor $e^{2}/\ell$ is set to
one in our units.

Let us now discuss an {\it effective} short-range interaction which
has been introduced for the Hall systems at filling fractions
$\nu=1/m$, say $1/3$ (Haldane's ``pseudo-potentials''
\halb \hala). We would like to motivate it by
using a renormalization group picture. Indeed, the
phenomenological and numerical results in the literature on
the Laughlin wave function strongly suggest such a picture.
We have in mind the renormalization-group flow in the space
of effective interactions \srange\ and we would like
to associate a fixed point to each $\nu=1/m$ plateau, with a
corresponding effective interaction.
The main supporting facts are:
\bigskip
\noindent
{\it i)} There is {\it universality}, namely the Laughlin wave
function
\eqn\efflwf{\phi_{\nu}(z_1,\dots,z_N)\ =\ \prod_{i<j}{(z_i - z_j)}
^{m},\qquad \nu={1 \over m},}
is an extremely good approximation to the numerical ground state for a
large class of repulsive interactions at $\nu=1/m$ \laugha.
The approximation
improves as one approaches the thermodynamical limit \ref\fano{
F. D. M. Haldane and E. H. Rezayi, {\it Phys. Rev. Lett.} {\bf 54}
(1985) 237; G. Fano, F. Ortolani and E. Colombo, {\it Phys. Rev.}
{\bf B 34} (1986) 2670.}
\footnote{*}{Compare with
finite-size effects at the critical point in field theory.}.
\bigskip
\noindent
{\it ii)} The Laughlin wave function is the {\it exact}
incompressible ground state for the short-range potential \hala\trukiv,
\eqn\srpot{H^{(m)}\ =\ \sum_{\scriptstyle k=1\atop\scriptstyle{k
\hbox{ odd}}}^{m-2}
e_{k} V_{k},}
where $e_{1},\dots,e_{m-2}$ are positive couplings
(this statement will be proven later).
\bigskip
\noindent
{\it iii)} There is a tunable mass scale $\Lambda$ in the
field-theoretical sense
\eqn\scale{\Lambda^{2} \sim \alpha \qquad \left( \sim {1\over B}
\qquad {\rm{ finite\ sample}} \right) ,}
which controls the density $\nu \propto 1/J \sim \alpha$, by
eqs. \nup\confj, and allows one to move among plateaus.
Indeed, the typical
inter-electron distance (in correct quantum units) is $1/\sqrt \alpha$
, which plays the role of correlation length.
\bigskip
\noindent
{\it iv)} The incompressibility of the $\nu=1/m$ ground-states
should be of the same nature as in the $\nu =1$ case.
\bigskip
These facts suggest the following renormalization group
scenario. The Coulomb interaction (including possible lattice
impurities) corresponds to the ``bare'', or classical, interaction,
whose couplings are given by
\coulv.
At a given value of $\alpha$, the electrons test
the Coulomb interaction at all scales smaller than $1/{\sqrt{\alpha}}$.
The bare couplings $e_k$, originally
monotonically decreasing for $k \to \infty$,
have a renormalization group flow.

For ($\alpha$ such that) $\nu \simeq 1/m$, we assume that this
flow is attracted to the fixed
point,
\eqn\fixpo{\{ e_{k}^{(eff)} ({\displaystyle{{1\over{\nu}}=m}}) \}=
\{ e_1 (m),\dots,e_{m-2} (m),0,0,\dots \},}
characterized by non-vanishing short-range effective couplings
($e_{k}$, $k< m$) and by vanishing long-range ones ($k \ge m$).
Intuitively, the latter
become irrelevant because they are not dynamically
important at the length scales allowed by the density.
Therefore, we are going to consider the short-range interaction
\srpot, as an
exact description of the interaction at the Hall plateaus $\nu=1/m$,
in the sense of fixed-points of the renormalization group.

Whilst this picture
needs to be clarified by a computation of beta functions, it is
substantiated by the numerical results reported in ref. \halb.
Haldane observed that the ground state and the lowest lying states
behave smoothly when the long-range couplings of the Coulomb interaction
$e_{k}^{(c)},\ k \ge m$, are turned off. Here we only
presented his arguments in a more appealing language, by stating
that the Coulomb and short-range $H^{(m)}$ interaction belong to
the same universality class for $\nu=1/m$.

Our characterization of the plateaus as fixed-points is
found for integer $\nu$ in the phase diagram of the
effective field theory approach of ref.
\ref\pruisk{A. M. M. Pruisken, {\it ``Field Theory, Scaling and
the Localization Problem''}, in \qhe .}.
Recently, an extension of
this phase diagram for fractional filling
has been proposed \lutros. The
$\nu=1/m$ plateaus appear as fixed points on the boundary of the
diagram, corresponding to accumulation points of a purely
massive phase, like the high-temperature fixed point of the
Ising model, where $\Lambda^2 = T -T_c \to \infty$.
\vfill 
\eject 
\noindent
{\bf 4.3 Exactness and infinite symmetry of Laughlin's wave
function}
\bigskip
Let us now study the properties of the fixed-point theory at
$\nu=1/m$. It describes electrons constrained to the lowest
Landau level and interacting via the two-body repulsive potential
\eqn\hmeq{H^{(m)}\ =\ \sum_{\scriptstyle k=1\atop\scriptstyle{k
\hbox{ odd}}}^{m-2} e_{k}  \ \sum_{i<j}
V_{k}(i,j)\ ,\qquad\ \  V_{k}(i,j)\ =\ |\ J_{ij} =k\
\rangle\ \langle J_{ij} =k\ |,}
where $V_{k}$ is the short-range potential defined previously.
A more explicit form of its action on
the holomorphic wave function is given by
\eqn\vkwf{\eqalign{V_{k}(i,j)\ &
\varphi(z_1, \dots,z_i,\dots,z_j,\dots,z_N)\ = \cr
(z_i & - z_j)^{k}\ {1\over k!} \left( {\partial \over{\partial x}}
\right)^{k} \varphi
\left.\left(z_1,\dots,{{z_i + z_j + x}\over 2},
    \dots,{{z_i + z_j - x}\over 2},\dots,z_N \right)
    \right\vert_{x=0} \cr}}
which generalizes the two-body case \vphipw.

By antisymmetry, the analytic wave-function $\varphi$ in \gpsilll\
has the general form
\eqn\gfphi{\varphi= \prod_{i<j} (z_i -z_j)\ P(z_1,\dots,z_N),}
where $\prod (z_i - z_j)$ is the Vandermonde determinant and $P$
is a completely symmetric homogeneous polynomial.

For the first non-trivial case, $\nu=1/3$, the Hamiltonian is
\eqn\hamsr{H^{(3)}\ =\ e_{1} \sum_{i<j} V_{1} (i,j) \qquad ,
\qquad e_{1} > 0 \ .}
We now discuss its spectrum.
The $E=0$ eigenspace is highly degenerate and contains wave-functions
which go to zero at coincidence points faster than $(z_i - z_j)$
for any pair $i,j=1,\dots,N$.
The $E>0$ eigenstates contain those which vanish as $O(z_i-z_j)$,
{\it i.e.} minimally.
For example, $H^{(3)}$ can be easily
diagonalized for $N=2$, by using eq. \vkwf,
\eqn\diagtwo{\eqalign{ E &= 0,\quad \phi_{0} = {(z_1 - z_2)}^{3}
\ P(z_1,z_2), \qquad {\Theta}_{0} = 1 - V_{1} , \cr
 E &= e_{1},\quad \phi_{1} = {(z_1 - z_2)}
\ Q(z_1+z_2), \qquad {\Theta}_{1} = e_1\ V_{1} , \cr }}
\noindent
where ${\Theta}_{0} , {\Theta}_{1}$ are the projectors on the
corresponding eigenspaces,
\eqn\thetas{{\bf 1} = {\Theta}_{0} + {\Theta}_{1} , \quad
{\Theta}_{0} {\Theta}_{1} =\ {\Theta}_{1} {\Theta}_{0} =\ 0,
\quad
H^{(3)} {\Theta}_{0} =\ 0, \quad
H^{(3)} {\Theta}_{1} =\ e_1 \ {\Theta}_{1} \quad .}

We now find the ground state of the interacting theory and later
discuss its incompressibility.
We prove the following statement:
{\it the $E=0$ eigenspace of the Hamiltonian \hamsr\ consists
of wave functions which contain the third power of the
Vandermonde as a factor.}
\eqn\eqnphio{\varphi_{0} = \prod_{i<j} {(z_i - z_j)}^{3} P(z_1,\dots,
z_N), \qquad\leftrightarrow \qquad H^{(3)} \varphi_{0} = 0,}
where $P$ is a symmetric homogeneous polynomial.
Clearly wave-functions of this form have vanishing energy, by
eq. \vkwf. Conversely, let us find the solutions to
$H^{(3)} \phi= 0$. Note that projectors have positive expectation
value, thus we can write
\eqn\pospro{0\ =\ \langle \phi| H^{(3)} | \phi \rangle = e_1 \ \sum_{i<j}
\|\ V_{1} (i,j)|\phi \rangle \|^{2}\ \to\ V_{1}(i,j) \phi = 0\ , \quad
\forall\ i,j .}
There are as many independent conditions as the number of pairs. Again,
by eq. \vkwf, $\phi$ should vanish faster than $(z_i - z_j)$
for any pair; thus, by antisymmetry, it should have at least a
third-order
zero for every pair. Moreover, $\phi$ is an analytic polynomial ({\it
i.e.}, entire) function of $z_1,\dots,z_N$, which is completely
determined
by its zeroes (whose number is given by $J$). Therefore, $\phi$
should have a third power of the Vandermonde determinant as a factor,
{\it q.e.d.}\footnote{*}{
The structure of inclusion of the Hilbert space for
$\nu=1, {1\over 3}, {1\over 5},\dots$
resembles the projections in rational conformal field theories,
like the projection of the free bosonic Fock space carried
on by the Coulomb gas technique \ref\felder{G. Felder, {\it Nucl. Phys.}
{\bf B 317} (1989) 215, {\hbox{ erratum}} {\it ibid.} {\bf 324} (1989)
548.}.}

A corollary to this statement is that the Laughlin wave function
\efflwf\ is the {\it exact
non-degenerate ground state} at $\nu=1/3$, as firstly observed
by Haldane \halb \hala.
On the plane, the value $\nu=1/3$ corresponds to
angular momentum $J=3N(N-1)/2$. Actually,
$J=\sum_{i} z_i \partial_i$ gives the degree of homogeneity, so
that $P$ has degree zero in this case, $P=1$; thus, $\varphi_{0}$
at $\nu=1/3$ is the Laughlin wave function \efflwf.
This ground state is unique and incompressible, because it
has the minimal value of $J$ in the $E=0$ eigenspace. Indeed,
in eq. \eqnphio\ we proved that $E=0$ implies
$J \ge 3N(N-1)/2$. Conversely, any
state of smaller $J$ should necessarily have $E > 0$, {\it i.e.},
it has an energy gap.

On the other hand, decompressions,
$J > 3N(N-1)/2$, also develop a gap after inclusion of the confining
potential \confj
\eqn\confff{H^{(3)} \rightarrow {\cal{G}}^{(3)} = H^{(3)}
+ \alpha {\cal{L}}_{0,0},}
in complete analogy with the integer filling case discussed in
section 3.

We shall now discuss the geometrical conditions expressing the
incompressibility of the Laughlin wave function.
To this end, we introduce the projector
$\Theta_{0}$ on the $H^{(3)}=0$ subspace
\eqn\thetres{\Theta_{0}\ =\ \sum_{i} |\ \varphi_{0,i} \rangle\ \langle
\varphi_{0,i}\ |, \qquad H^{(3)} \Theta_{0} = 0,}
\noindent
where $\{ \varphi_{0,i} \}$ is an orthogonal set of $\varphi_{0}$'s
of eq. \diagtwo.

Next, we modify the generators ${\cal L}_{n,m}$ of the non-interacting
symmetry algebra \defl\ as follows
\eqn\lnmpr{{\cal{L}}_{n,m} = \sum_{i} z_{i}^{n+1}\ \partial_
{i}^{m+1}, \qquad\rightarrow\qquad
{\hat{\cal L}}_{n,m}\ \equiv\ \Theta_{0} \ {\cal{L}}_{n,m}\  \Theta_{0}.}
The projected generators $\hat{\cal{L}}_{n,m}$ commute with
$H^{(3)}$ and produce compressions and decompressions in the
$E=0$ subspace described before \eqnphio.
They satisfy a modified algebra, due to the projection
$\Theta_0$. Nevertheless, their angular momentum eigenvalue
is unchanged because ${\cal{L}}_{0,0}={\hat{\cal{L}}}_{0,0}$, and
$[{\hat{\cal L}}_{0,0},{\hat{\cal L}}_{n,m}]=(n-m) {\hat{\cal L}}
_{n,m}$, in the $E=0$ subspace.

Acting on the Laughlin wave function \efflwf, we find
\eqn\lnmlwf{
{\hat{\cal L}}_{n,m}\ \phi_{\scriptstyle{1\over 3}} \ = \
\Theta_{0}\ {\cal{L}}_{n,m}\  \Theta_{0}\ \prod_{i<j}
{(z_i - z_j)}^{3}\ =\ 0, \quad {\hbox{for}}\quad  n<m,\ m\ge 0.}
\noindent
Proof: consider $\Phi = {\cal{L}}_{n,m} \Theta_{0} \prod
{(z_i - z_j)}^{3}$. ${\cal{L}}_{n,m}$ lowers the angular
momentum, thus $\Phi$ has
$J=3N(N-1)/2\ -(m-n)\ <\ 3N(N-1)/2\ \ $.
We already proved that functions
with such values of $J$ are necessarily a superposition of
$E > 0$ eigenstates only. Therefore, $\Theta_{0} \Phi=0$, by eq.\thetres,
{\it q.e.d.}.

Therefore, {\it eqs. \lnmlwf\ are the infinite incompressibility
conditions satisfied by the Laughlin wave function at}
$\nu=1/m$.

We achieve a description of incompressibility of the quantum fluid
analogous to the integer $\nu$ case. Among the transitions generated
by the ${\cal L}_{n,m}$, we neglected compressions to $E>0$ states,
which are forbidden by the macroscopic gap. The remaining ones
(the ${\hat{\cal L}}_{n,m}$, $n<m, m\ge 0$), which are energetically
allowed,
were used to establish the incompressibility of the ground state,
{\it i.e.}, its infinite symmetry.
Notice that this exact characterization
is possible due to the peculiar properties of the two-body effective
interaction, which renders the interacting theory much like the free one.

For the sake of completeness, a remark may be useful at this point.
We treated differently compressions and decompressions both in the
integer and fractional cases. Following Haldane \halb, we gave a
special role to compressions, because they characterize the effective
interaction which develops at the plateaus. Indeed, being a short-range
interaction, it allows for any decompression. Therefore, compressions
are characteristic, and determine the relevant symmetry algebra,
while decompressions are generic.

Unfortunately, at present we lack a complete understanding of
the projection $\Theta_0$ \thetas, and thus, of the
${\hat{\cal{L}}}_{n,m}$ algebra in this more interesting
fractional case. Clearly, a second-quantized formulation will
make manifest the action of the ${\hat{\cal L}}_{n,n}$ in the
$E=0$ subspace.

Nevertheless, it is interesting to remark that
the conformal subalgebra is preserved by the
projection,
$\ {\hat{\cal{L}}}_{n,0} = {\cal{L}}_{n,0}$.
Indeed, these transformations are locally a rotation and a
dilatation, which keep the order of the zeroes of the
wave function. This may be an indication that this subalgebra
plays a distinguished role at each one of the plateaus.
\vfill 
\eject 
\newsec{Summary and perspectives}
\bigskip
In summary, we have exposed the existence of an infinite symmetry
in the many-body quantum problem of planar
electrons in a uniform magnetic field. In the absence of two-body
interactions, the symmetry generators satisfy an algebra of (quantum)
area-preserving diffeomorphisms. Its classical origin are
canonical transformations of the four-dimensional phase space
that leave invariant the Hamiltonian. Upon quantization, the external
magnetic field quenches the kinetic energy and the phase space is
consequently reduced from four to two dimensions.

At the quantum level, these generators have been used to describe
excitations within any Landau level. The incompressibility of
completely filled levels was expressed by a set of highest weight
conditions \invac ,\vacinv, of the type which appear in conformal
field theory.

In the presence of Haldane's effective two-body interactions,
the basic picture does not change: there is still an infinite
algebra and the Laughlin ground states \lauwf\
are annihlilated by an infinite subset of generators.
In this case, the
explicit expressions of the generators and their algebra are not
fully understood at the moment.
The exact nature of the latter and the
eventual emergence of a full Virasoro algebra
constitute important unsettled
issues which are under current investigation.

Another interesting problem is to further elucidate the geometrical
properties of excited states. For example, let us consider the
(second-quantized) Fourier-transformed density operator $\rho(
{\bf k})$:
\eqn\eqrho{\rho ({\bf k}) = \int d^{2} {\bf k}\  e^{\displaystyle
i {\bf k \cdot x} }\ \rho({\bf x}). }
When projected onto the first Landau level this operator can be
written,
\eqn\rholnm{\rho ( {\bf k}) = e^{-{1\over 4} \displaystyle{
k {\bar k}}}
\sum_{n,m} {1\over{n!m!}} \left( {i {\bar k} \over{2}} \right )
^{n} \left( {i  k \over{2}} \right )^{m} {\cal{L}}_{n-1,m-1}.}
Due to the character of the ${\cal{L}}_{n,m}$, it
creates intra-level bosonic excitations labelled by
a wave vector ${\bf k}$: indeed, these are the
{\it magneto-phonons} and {\it magneto-rotons} neutral excitations
studied in
\ref\girv{S. M. Girvin, {\it ``Collective Excitations''}, in \qhe;
S. M. Girvin, A. H. MacDonald and P. M. Platzman, {\it
Phys. Rev.} {\bf B 33} (1986) 2481.}.

More relevant for Laughlin's theory are the
{\it quasi-hole} and {\it quasi-particle} charged excitations.
We expect these to fall into representations of the projected
algebra of the fractional case. Hopefully,
this will provide the correct characterization of anyons as
``vertex operators''.

As a last point, we mention the geometrical
treatment of the Fermi fluid in terms of {\it distribution
functions}. The expectation value of the density operator
$\rho ( {\bf x},t)$ in any quantum state gives a
corresponding quantum distribution
function $w({\bf x},t)$ of the fluid. This expectation value
is also expressed in terms of ${\cal{L}}_{n,m}$
\eqn\wcomp{w({\bf x},t)= {1\over \pi}\ e^{\displaystyle{- z
{\bar z} }}\
\sum_{n,m} {1\over{n!m!}} z^{n} {\bar z}^{m}
\langle {\cal{L}}_{n-1,m-1}\rangle.}
Thus, the space of all distribution functions $w$ carries a
representation of $W_{\infty}$. The semi-classical approximation
consists in considering distribution functions which
carry a representation of $w_{\infty}$, which is a more
tractable algebra.
Using this framework, we hope to understand better its geometrical
meaning in terms of deformations of the droplet of fluid and the
edge states \ref\edge{X. G. Wen, {\it Phys. Rev. Lett.} {\bf 64}
(1990) 2206; M. Stone, {\it Ann. Phys.} (NY) {\bf 207} (1991) 38;
J. Fr\"ohlich and T. Kerler, {\it Nucl. Phys.} {\bf B 354} (1991)
365.}.
This semiclassical construction was recently discussed
in a ``string inspired''
geometrical treatment of $(1+1)$-dimensional Fermi fluids
\ref\wfer{S. Iso, D. Karabali and B. Sakita, {\it ``One dimensional
fermions as two-dimensional droplets via Chern-Simons theory''},
CCNY-HEP-92/1 preprint; A. Dhar, G. Mandal and S. R. Wadia, {\it ``
Classical
Fermi fluid and geometrical action for c=1''}, IASSNS-HEP-91/89
preprint.}.
Note, however, that the Fermi
fluid considered here is $(2+1)$-dimensional and that the reduction
to a two-dimensional phase space is produced by the
external magnetic field.
\bigskip
\bigskip
\noindent
{\bf Acknowledgments}
\bigskip
\noindent
We gratefully acknowledge helpful discussions with Luis Alvarez-Gaum\'e
and Andy L\"utken.
We also thank Gerald Dunne for useful comments on the manuscript.

\listrefs
\end